\documentclass[amsmath,amssymb,aps,prd,11pt,tightenlines,superscriptaddress,nofootinbib,preprintnumbers,notitlepage]{revtex4-1}

\usepackage{graphicx}
\usepackage{tabularx}
\usepackage{mathtools}
\usepackage{url}
\usepackage{hyperref}
\usepackage{color,xcolor}
\usepackage{feynmp}
\makeatletter
\@ifundefined{pdfoutput}{}{\DeclareGraphicsRule{*}{mps}{*}{}}
\makeatother

\newcommand{\arxiv}[1]{\href{http://arxiv.org/abs/#1}{arXiv:#1}}
\newcommand{\arXiv}[1]{\href{http://arxiv.org/abs/#1}{arXiv:#1}}
\newcommand{\fermilat}{{\it Fermi}--LAT }
\newcommand{\igroi}{$15^\circ \times 15^\circ$}

\newcommand\one{\leavevmode\hbox{\small1\normalsize\kern-.33em1}}

\newcommand{\qqqquad}{\qquad \qquad \qquad}

\newcommand{\abs}[1]{\left |#1\right |}

\providecommand{\nni}{\tilde{\chi}_i^0}
\providecommand{\nnj}{\tilde{\chi}_j^0}
\providecommand{\nne}{\tilde{\chi}_1^0}

\providecommand{\cpe}{\tilde{\chi}_1^+}

\providecommand{\cme}{\tilde{\chi}_1^-}

\providecommand{\cpmj}{\tilde{\chi}_j^\pm}
\providecommand{\cpme}{\tilde{\chi}_1^\pm}

\newcommand{\mch}[1]{m_{\tilde{\chi}^+_{#1}}}
\newcommand{\mne}[1]{m_{\tilde{\chi}^0_{#1}}}

\newcommand{\gev}{\text{GeV}}
\newcommand{\tev}{\text{TeV}}

\newcommand{\br}{\text{BR}}

\def\slashchar#1{\setbox0=\hbox{$#1$}           
   \dimen0=\wd0                                 
   \setbox1=\hbox{/} \dimen1=\wd1               
   \ifdim\dimen0>\dimen1                        
      \rlap{\hbox to \dimen0{\hfil/\hfil}}      
      #1                                        
   \else                                        
      \rlap{\hbox to \dimen1{\hfil$#1$\hfil}}   
      /                                         
   \fi}

\newcommand{\eg}{\textsl{e.g.}\;}

\newcommand{\be}{\begin{eqnarray*}}
\newcommand{\ee}{\end{eqnarray*}}

\newcommand{\bee}{\begin{eqnarray}}
\newcommand{\eee}{\end{eqnarray}}
\newcommand{\beeq}{\begin{equation}}
\newcommand{\eeeq}{\end{equation}}

\begin{document}

\title{Saving the MSSM from the Galactic Center Excess}

\author{Anja Butter}
\affiliation{Institut f\"ur Theoretische Physik, Universit\"at Heidelberg, Germany}

\author{Simona Murgia}
\affiliation{Department of Physics and Astronomy, University of California, Irvine, CA 92697, USA}

\author{Tilman Plehn}
\affiliation{Institut f\"ur Theoretische Physik, Universit\"at Heidelberg, Germany}

\author{Tim M.P. Tait}
\affiliation{Department of Physics and Astronomy, University of California, Irvine, CA 92697, USA}

\preprint{UCI-HEP-TR-2016-23}

\begin{abstract}
  The minimal supersymmetric setup offers a comprehensive framework to
  interpret the \fermilat Galactic center excess.  Taking into account
  experimental, theoretical, and astrophysical uncertainties we can
  identify valid parameter regions linked to different annihilation
  channels. They extend to dark matter masses above 250~GeV. There
  exists a very mild tension between the observed relic density and the
  annihilation rate in the center of our galaxy for specific channels.
  The strongest additional constraints come from the new generation of
  direct detection experiments, ruling out much of the light and
  intermediate dark matter mass regime and giving preference to
  heavier dark matter annihilating into a pair of top quarks.
\end{abstract}

\maketitle
\tableofcontents

\begin{fmffile}{feyn}
\newpage
\section{Introduction}
\label{sec:intro}

While the existence of an unknown dark matter as the primary matter
component of today's Universe is solidly established, its particle
nature remains elusive.  A broad experimental program seeks to shed
light on this question by searching for dark matter indirectly through
the products of its annihilation, directly scattering with terrestrial
targets, or being produced at colliders.  Among indirect searches,
gamma rays with GeV-range energies are a particularly effective
messenger because they propagate unhindered on galactic scales, and
thus can be effectively traced back along the direction of their
origin.  In recent years, the {\it Fermi} Large Area Telescope ({\it
  Fermi}-LAT) has mapped out the gamma-ray sky with unprecedented
resolution, leading to the current best limits on the annihilation
cross section for dark matter particles with masses around 100~GeV.\bigskip

Remarkably, the {\it Fermi}-LAT data contains an indication of what
appears to be an excess of gamma rays from the direction of the
Galactic center (GC) above the predictions from astrophysical models,
with spatial morphology and spectrum consistent with expectations  for
the annihilation of a thermal
relic~\cite{Goodenough:2009gk,hooperon2,weniger_tale}.  The \fermilat
Collaboration has released its own
analysis~\cite{TheFermi-LAT:2015kwa} of the gamma rays from the
direction of the GC based on specialized interstellar emission models
(IEMs). These models allow for a determination of the gamma-ray fore/background originating from cosmic rays interacting with the
interstellar gas and radiation field, and for a separation from the
contribution from within roughly 1~kpc of the GC along the line of
sight toward it. The GC excess persists in this analysis, and its
spectral properties display a strong dependence on the assumed IEM,
making it challenging to conclusively identify its origin.  It thus
remains unclear whether this signal arises from dark matter
annihilation rather than from other, more mundane sources, such as a
population of unresolved millisecond pulsars, cosmic-ray proton or
electron outbursts, additional cosmic ray sources, and/or emission
from a stellar over-density in the Galactic bulge~\cite{stuff}.  An
interesting development is the use of statistical tools which indicate
that the  excess   displays more clustering than would be
expected from Poisson noise from smooth components~\cite{sources}.
However, it remains difficult with the current models to disentangle
whether this feature represents a property of the excess itself, or
un-modeled variation in the background
components~\cite{Horiuchi:2016zwu}.\bigskip

While it is premature to claim that the GC excess represents a
confirmed signal of dark matter annihilation, 
in this paper we interpret its properties under the assumption that it does in the framework of the minimal supersymmetric
extension of the Standard Model (MSSM).  The MSSM is a prototypical
model of weakly interacting massive particles.  In the region of
parameter space for which the lightest supersymmetric particle (LSP)
is a neutralino, a rich vision for dark matter emerges, largely
dictated by its component fractions of electroweak singlet, doublet,
and triplet representations~\cite{wimp}.  Despite this flexibility,
early investigations found it challenging to fit the original
characterizations of the GC excess in the MSSM~\cite{hooperon_scan}
due to the generic requirement of efficient
mediators~\cite{hooperon_simplified} naturally present in extended
models such as the
NMSSM~\cite{hooperon_extendedmssm,hooperon_nmssm,hooperon_sfitter}.

In this article, we perform the first analysis of the MSSM parameter
space capable of describing the GC excess as extracted by the
\fermilat Collaboration in Ref.~\cite{TheFermi-LAT:2015kwa}, including
the range of spectra corresponding to the full suite of models for the
interstellar emission developed therein.  We examine how this wide
range of spectra opens up regions of the MSSM parameter space
describing the excess~\cite{hooperon_mssm} by performing global fits
to these spectra in the \textsc{SFitter} framework~\cite{sfitter},
consistently with the thermal relic density, the light Higgs boson
mass, and the standard set of low energy indirect constraints.  The
power of such a global analysis rests in its ability to interpret the
wide range of relevant experimental
observations~\cite{sfitter_planck,hooperon_fit,weniger_mssm,fittino}.

\section{The Galactic center excess}
\label{sec:gce}

\begin{figure}[t]
\includegraphics[width=0.65\textwidth]{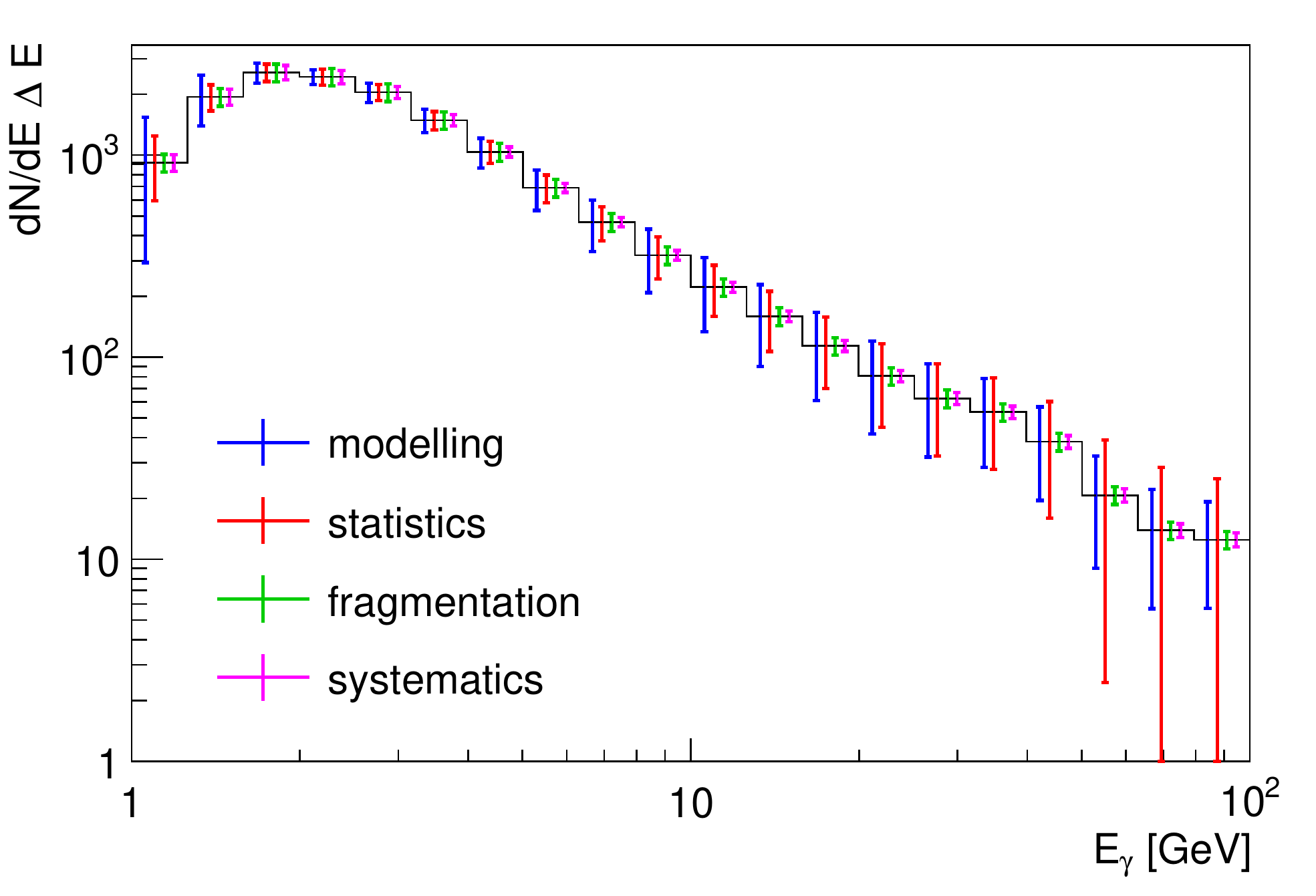}
\caption{GC excess spectrum from~\cite{TheFermi-LAT:2015kwa},
  including uncertainties from the interstellar emission model and
  fragmentation, as well as instrumental systematics and statistical
  uncertainties, as described in the text.}
\label{fig:spectrum}
\end{figure}

The \fermilat Collaboration determination of the GC
excess is based on the first 62 months of
data in a \igroi\ region in the direction of the GC in the energy
range $E_\gamma = 1~...~100$~GeV.  In order to minimize the bias from the data
toward the GC, the methodology developed
in Ref.~\cite{TheFermi-LAT:2015kwa} employs regions outside of the
\igroi\ region for the determination of the fore/background emission.
Furthermore, the point sources are determined self-consistently
together with each IEM. This is a crucial improvement over previous
analyses, as the determination of the point sources in this region is
strongly dependent on the IEM.  We refer the reader
to Ref.~\cite{TheFermi-LAT:2015kwa} for a more detailed description of
these models and their associated point sources.\bigskip

We adopt the \fermilat GC excess
spectrum for a spectral model assumed to
be a power-law function in each of 10 energy bands, equally spaced in
logarithmic energy over $E_\gamma = 1~...~100$~GeV, shown in Fig.~\ref{fig:spectrum}. The obtained
spectral envelope spans the full set of IEMs and therefore encompasses
the interstellar emission modeling
uncertainty from Ref.~\cite{TheFermi-LAT:2015kwa}, uncorrelated bin-by-bin in
the energy spectrum. Unlike a correlated global modification, this
allows for a more sizable change in the shape of the photon spectrum.
The exclusive log-likelihood is flat within the envelope, in harmony
with the assumption for theoretical uncertainties in
\textsc{SFitter}~\cite{sfitter}. Combined with a profile likelihood
this is equivalent to using the \textsc{RFit} scheme~\cite{rfit}.  In
addition to the modeling uncertainty on the interstellar emission,
which is the dominant source of uncertainty over most of the energy
range, we include the statistical error on the signal rate after
background subtraction. The statistical uncertainty thus reflects the
combined statistical uncertainty of both of signal and
background~\cite{sfitter}, and it is uncorrelated between bins. Note
that this is the leading uncertainty for $E_\gamma \gtrsim 40$~GeV.
Furthermore, we include a 10\% uncertainty from the fragmentation function for
photons~\cite{weniger_mssm},  treated as un-correlated between
energy bins and Gaussian distributed.  Finally, we include the
systematic error on the \fermilat effective area~\cite{Ackermann:2012kna},   treated as fully correlated between bins and also Gaussian distributed.

The primary observables for the GC excess are  the annihilation
cross section, which characterizes the over-all brightness of the
excess, and its spectral shape binned in energy.  The annihilation
cross section itself is fully degenerate with the $J$-factor, which
quantifies the integral of the square of the dark matter density along the line
of sight encompassed within the $15^\circ \times 15^\circ$ region
employed to extract the signal in Ref.~\cite{TheFermi-LAT:2015kwa}.  The
best estimates for the uncertainty in the $J$-factor are that it can
vary by roughly a factor of two in the region of interest~\cite{Abazajian:2015raa}.

\section{MSSM annihilation channels}
\label{sec:channels}

Our MSSM parameter analysis can be most easily organized in terms of
the the dominant dark matter annihilation channels.  For a typical
weakly interacting dark matter candidate comprising all of the dark
matter and following a standard cosmological history, the same
annihilation cross section which explains the GC excess also
determines the thermal relic abundance.  However, in a theory
containing multiple components of dark matter and/or a nonstandard
cosmology, the relic abundance and the annihilation cross section are
less correlated.  For this reason, in this section we remain somewhat
agnostic as to whether the dark matter abundance arises from the usual
freeze-out calculation, whereas in Sec.~\ref{sec:global} we fit both
the GC excess and the relic abundance assuming a standard cosmological
history.

We focus on the most important MSSM parameters determining the dark
matter properties: the wino mass $M_2$, the higgsino mass parameter
$\mu$, and the bino mass $M_1$.  As we will see below, the masses of
the heavy Higgs states $m_{A,H}$ can play an important role for dark
matter annihilation.  The light Higgs mass $m_h$ is adjusted with the
help of $\tan \beta$, $A_t$. The third-generation squark masses, with
the remaining sleptons, squarks, and gluinos are assumed to decouple,
as suggested by the Higgs mass and the direct limits from the null
results of LHC searches. For all scenarios we require the light Higgs
mass to match the measured value $m_h = 126$~GeV and charginos to
be heavier than the LEP limit of 103~GeV.

As for all \textsc{SFitter} analyses~\cite{sfitter} we calculate the
MSSM spectrum with \textsc{SuSpect3}~\cite{suspect}, while the Higgs
branching ratios are computed using \textsc{Susy-Hit} and
\textsc{HDecay}~\cite{s-hit}. The relic density and the indirect
annihilation rate are calculated with
\textsc{MicrOMEGAs}~\cite{micromegas}.\bigskip

\begin{figure}[b!]
\begin{fmfgraph*}(80,60)
\fmfset{arrow_len}{2mm}
\fmfleft{i1,i2}
\fmfright{o1,o2}
\fmf{plain,tension=0.4,width=0.6}{i2,v1}
\fmf{photon,tension=0.4,width=0.6}{i2,v1}
\fmf{plain,tension=0.4,width=0.6}{v1,i1}
\fmf{photon,tension=0.4,width=0.6}{v1,i1}
\fmf{dashes,tension=0.4,label=$h^0$}{v1,v2}
\fmf{fermion,tension=0.6,width=0.6}{v2,o1}
\fmf{fermion,tension=0.6,width=0.6}{o2,v2}
\fmflabel{$\nne$}{i1}
\fmflabel{$\nne$}{i2}
\fmflabel{$b$}{o1}
\fmflabel{$b$}{o2}
\end{fmfgraph*}
\hspace*{7mm}
\begin{fmfgraph*}(80,60)
\fmfset{arrow_len}{2mm}
\fmfleft{i1,i2}
\fmfright{o1,o2}
\fmf{plain,tension=0.4,width=0.6}{i2,v1}
\fmf{photon,tension=0.4,width=0.6}{i2,v1}
\fmf{plain,tension=0.4,width=0.6}{v2,i1}
\fmf{photon,tension=0.4,width=0.6}{v2,i1}
\fmf{plain,tension=0.2,label=$\cpmj$}{v1,v2}
\fmf{photon,tension=0.2}{v1,v2}
\fmf{photon,tension=0.7,width=0.6}{v2,o1}
\fmf{photon,tension=0.7,width=0.6}{o2,v1}
\fmflabel{$\nne$}{i1}
\fmflabel{$\nne$}{i2}
\fmflabel{$W^+$}{o1}
\fmflabel{$W^-$}{o2}
\end{fmfgraph*}
\hspace*{12mm}
\begin{fmfgraph*}(80,60)
\fmfset{arrow_len}{2mm}
\fmfleft{i1,i2}
\fmfright{o1,o2}
\fmf{plain,tension=0.4,width=0.6}{i2,v1}
\fmf{photon,tension=0.4,width=0.6}{i2,v1}
\fmf{plain,tension=0.4,width=0.6}{v2,i1}
\fmf{photon,tension=0.4,width=0.6}{v2,i1}
\fmf{plain,tension=0.2,label=$\nnj$}{v1,v2}
\fmf{photon,tension=0.2}{v1,v2}
\fmf{dashes,tension=0.7,width=0.6}{v2,o1}
\fmf{dashes,tension=0.7,width=0.6}{o2,v1}
\fmflabel{$\nne$}{i1}
\fmflabel{$\nne$}{i2}
\fmflabel{$h^0 Z$}{o1}
\fmflabel{$h^0 Z$}{o2}
\end{fmfgraph*}
\hspace*{12mm}
\begin{fmfgraph*}(80,60)
\fmfset{arrow_len}{2mm}
\fmfleft{i1,i2}
\fmfright{o1,o2}
\fmf{plain,tension=0.4,width=0.6}{i2,v1}
\fmf{photon,tension=0.4,width=0.6}{i2,v1}
\fmf{plain,tension=0.4,width=0.6}{v1,i1}
\fmf{photon,tension=0.4,width=0.6}{v1,i1}
\fmf{dashes,tension=0.4,label=$A^0 H^0$}{v1,v2}
\fmf{fermion,tension=0.6,width=0.6}{v2,o1}
\fmf{fermion,tension=0.6,width=0.6}{o2,v2}
\fmflabel{$\nne$}{i1}
\fmflabel{$\nne$}{i2}
\fmflabel{$t$}{o1}
\fmflabel{$t$}{o2}
\end{fmfgraph*}
\vspace*{5mm}
\caption{Feynman diagrams illustrating dark matter
  annihilation $\nne \nne \to b\bar{b}, WW, ZZ, hh,
  t\bar{t}$ in the MSSM.}
\label{fig:feyn}
\end{figure}
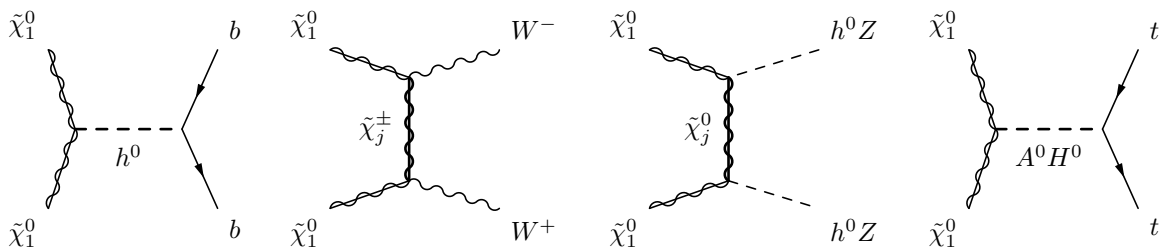

Representative Feynman diagrams for the most important annihilation
processes are shown in Figure~\ref{fig:feyn}.  Generically, it is
difficult to realize large enough cross sections to explain the GC
excess~\cite{sfitter_planck,jay,DMpapers} for an LSP with a suitable
mass.  For example, $t$-channel annihilation channels are generally
not very efficient and decouple rapidly with the mediator
mass~\cite{dm_eft}.  Large enough annihilation cross sections
typically occur for:
\begin{itemize}
\item $t$-channel chargino exchange driven by the coupling to
  $W$-bosons in the final state,
\begin{align}
g_{W \nne \cpe} =
\frac{g \sin \theta_w}{\cos \theta_w} \; 
\left( \frac{1}{\sqrt{2}} N_{14} V_{12}^* - N_{12} V_{11}^* \right) \; ,
\label{eq:w_coup}
\end{align}
  which is most efficient for charginos just above the LEP limit
  $\mch{1} = 103$~GeV.  A substantial coupling to $W$ bosons requires
  that the LSP contains a sizable fraction of either a wino higgsino
  fraction.
\item $t$-channel neutralino exchange, leading to $\nne \nne \to ZZ$
  or $\nne \nne \to hh$~\cite{weniger_tale}. For the former, the
  relevant coupling is an axial-vector coupling with strength
\begin{align}
g_{Z \nne \nni}
=\dfrac{g}{2 \cos \theta_w}
 \left( N_{13}N_{i3}-N_{14}N_{i4}\right) \; ,
\label{eq:z_coup}
\end{align}
  driven by the higgsino content. For the latter process, the relevant
  couplings are products of higgsino and gaugino fractions, requiring
  that the LSP be a highly mixed state,
\begin{align}
g_{h \nne \nni}
= \left( g N_{11} - g' N_{12} \right) \; 
   \left( \sin \alpha \; N_{13} + \cos \alpha \; N_{14} \right) \; .
\label{eq:h_coup}
\end{align}
  The mixing angle $\alpha$ describes the rotation of the scalar
  Higgses into mass eigenstates.
\item $t$-channel sfermion exchange, \eg tau sleptons. In this case,
  significant coupling requires a large wino fraction, which typically
  leads to excessively large annihilation into $W$ bosons for LSP
  masses below around 1~TeV.
\end{itemize}
More efficient are $s$-channel annihilation processes, particularly
when the the masses of the dark matter and the mediating particle are
arranged such that the annihilation benefits from the on-shell
resonance.  Candidates for $s$-channel mediators in the MSSM are:
\begin{itemize}
\item Vector $Z$-funnel annihilation through the Higgsino component,
  as illustrated in Eq.\eqref{eq:z_coup}. The coupling vanishes in the
  limit $\tan \beta \rightarrow 1$, due to approximately equal
  Higgsino fractions. Large $\tan \beta$ also reduces the predicted
  spin-independent direct detection cross section and therefore allows
  for a larger allowed parameter space.  Because the axial-vector
  component does not have a velocity suppression, the annihilation
  rate $\langle \sigma v \rangle$ usually prefers LSP masses slightly
  above or below 45~GeV; directly on the $Z$-pole the annihilation is
  too efficient.
\item Scalar $h$-funnel annihilation, where the LSP mass should be
  around $\mne{1} = 63$~GeV, slightly away from the resonance. The
  coupling in Eq.\eqref{eq:h_coup} relies on higgsino-gaugino mixing.
  Almost the entire neutralino annihilation rate through the light
  Higgs funnel goes to $b\bar{b}$, with small contributions or $\tau^+
  \tau^-$ and $WW$.
\item Heavy (pseudo-)scalar Higgs funnel annihilation, where the
  pseudo-scalar $A^0$ leads to an efficient $s$-wave annihilation. The coupling is again driven by higgsino-gaugino
  mixing. Heavy scalar decays to down-type fermions are enhanced by
  $\tan \beta$, which implies that for $\tan \beta \gtrsim 30$ the
  resonance pole structure of the $A$-funnel gets significantly
  washed out and a $b\bar{b}$ final state appears from this topology.
\end{itemize}
Finally, co-annihilation channels are an efficient means to realize
the relic density when there is an additional supersymmetric particle
within about 10\% of the LSP
mass~\cite{stau-co-annihilation,char-co-annihilation,stop-co-annihilation}.
For the light dark matter particles, usually associated with the
\fermilat GC excess, additional light charginos or sfermions are
strongly disfavored for example by LEP
constraints~\cite{lep_constraints}. For heavier dark matter,
co-annihilation can significantly contribute for example for processes
with a light chargino in the $t$-channel.\bigskip

The above annihilation mechanisms are often closely linked to LHC
search channels. For instance, $t$-channel chargino annihilation or
neutralino/chargino co-annihilation point to more than one light
electroweakino, where at least one of the additional light states is a
chargino. In this situation one can search for $\nnj \cpme$ or $\cpe
\cme$ production. One of the classic signatures are tri-leptons, which
become challenging when the mass differences between the chargino and
the neutralino become small~\cite{final_word}.  Similarly, $t$-channel
sfermion exchange or sfermion co-annihilation point towards another
light particle, which can be pair-produced through its QED or QCD
interactions. As long as the mass difference is not extremely small,
such light sfermions are accessible at the LHC, particularly when
colored.  The situation becomes more challenging when the mediator is
a Standard Model particle. To establish this mediator role one would
need to establish $Z$ or Higgs coupling to the dark matter sector, for
example through invisible $Z$~\cite{z_inv} and/or Higgs
decays~\cite{h_inv}.

\subsection*{$\chi \chi \to b\bar{b}$}

\begin{figure}[t]
\includegraphics[width=0.495\textwidth]{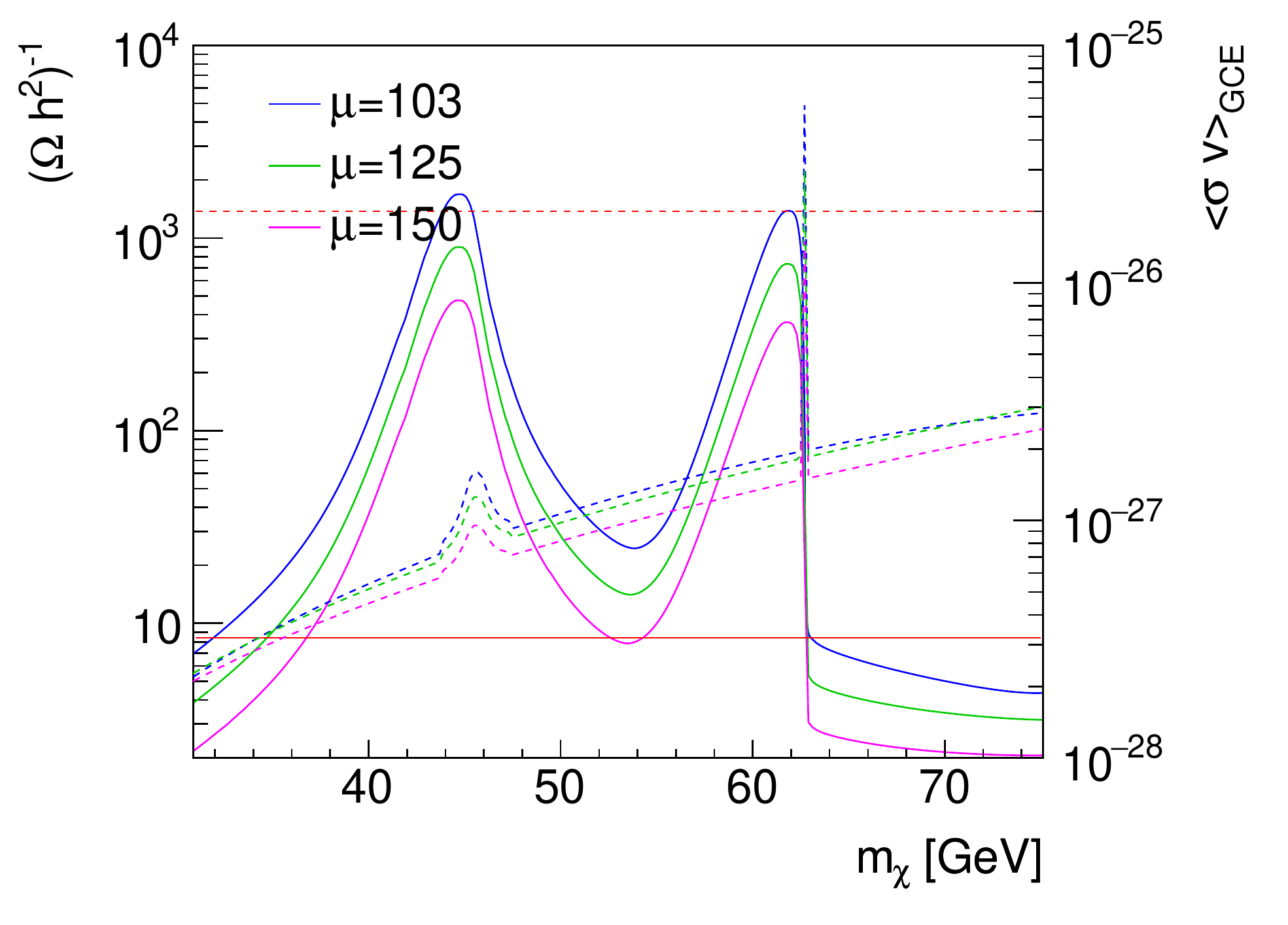}
\includegraphics[width=0.495\textwidth]{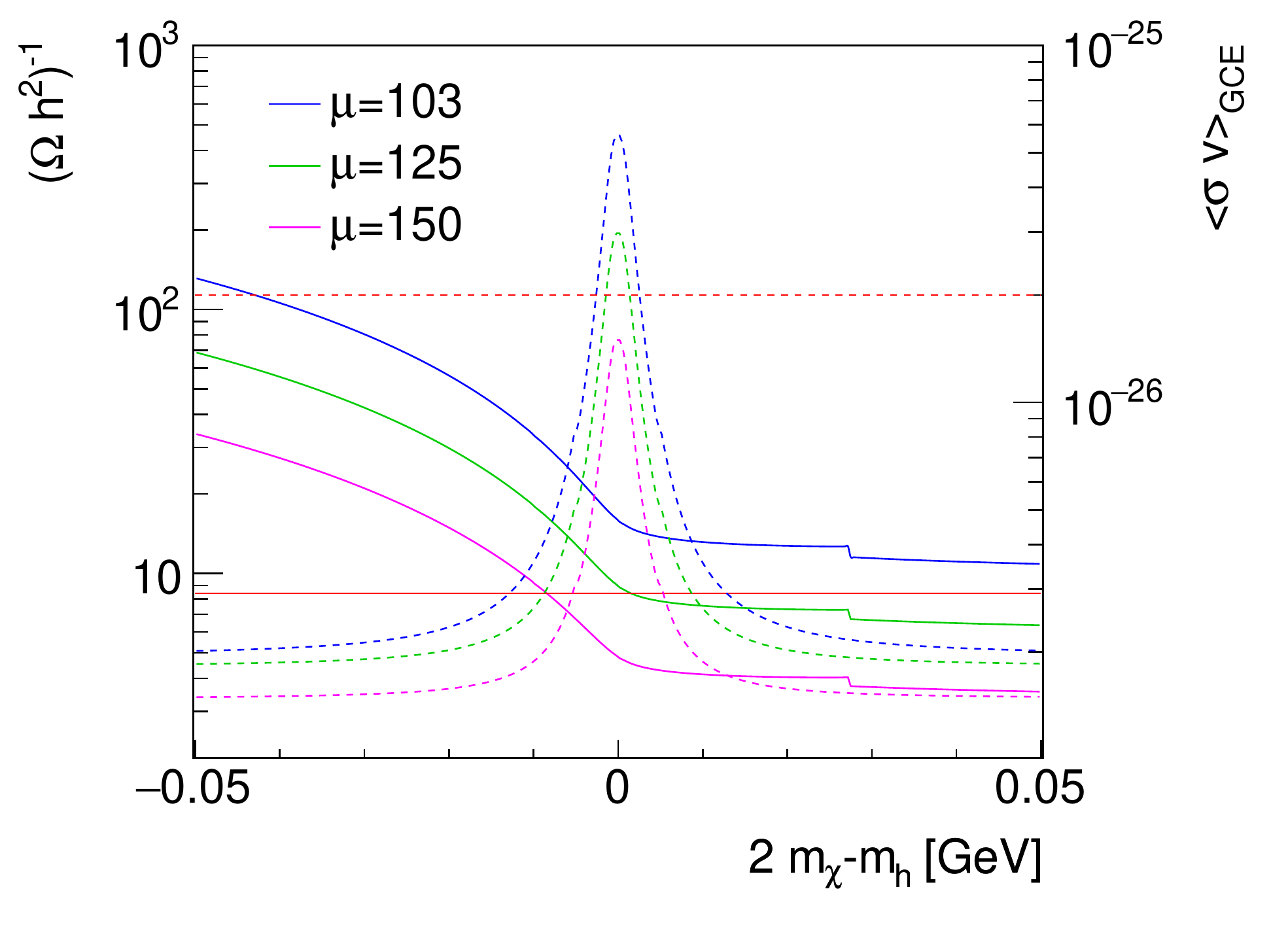}
\caption{Inverse relic density (dashed, left axis) and annihilation
  rate in the GC (solid, right axis) for an MSSM parameter point where
  the annihilation is dominated by $\nne \nne \to b\bar{b}$. The right
  panel is zoomed into the Higgs pole region. Additional model
  parameters are $\tan\beta=45$, and third-generation squark masses
  range around 1~TeV.}
\label{fig:bb}
\end{figure}

To define an MSSM scenario with a light neutralino responsible for the
GC excess we examine the regions of MSSM parameter space where the
annihilation $\nne \nne \to b\bar{b}$ dominates the dark matter
annihilation in Fig.~\ref{fig:bb}.  For light neutralinos,
annihilation tends to be dominated by the $s$-channel light Higgs
funnel, rather than the broad $A$-induced band.  The lightest
neutralino is mostly bino, with some higgsino content to couple to the
$Z$ and the light Higgs mediators, and negligible wino content ($M_2 =
700$~GeV). We also fix $\tan \beta = 45$, though the results are
rather insensitive to this choice.  The varying neutralino mass on the
$x$-axis is generated by adjusting $M_1$ for each of the fixed values
of $\mu$.

On the left $y$-axis in the left panel of Fig.~\ref{fig:bb} we show
the inverse relic density, proportional to the annihilation rate in
the early universe. The corresponding solid curves exhibit two
distinct peaks, one for $Z$-funnel annihilation and one for $h$-funnel
annihilation. For both peaks the width is given by the velocity
spectrum rather than the intrinsic width of the mediators. 
The enhancement of the two peaks over the continuum end up being
comparable, with the $Z$-funnel coupled to the axial-vector
current which is velocity suppressed with $v \lesssim 1/10$, whereas the
Higgs funnel is suppressed by the small bottom Yukawa coupling. The measured
relic density can be reproduced on the shoulders of the resonance
peaks, with a slight preference for larger $\mu$-values and hence
smaller couplings.

On the right $y$-axis of Fig.~\ref{fig:bb} (corresponding to the
dashed curves) we show the annihilation rate in the GC, with the rough
target rate indicated by the horizontal line. Because of the much
smaller velocities, the widths of the resonance peaks are now
determined by the physical widths of the $Z$ and the Higgs. The Higgs
resonance leads to much larger peak rates, because of the stronger
velocity suppression of the axial-vector coupling to the
$Z$-mediator. We observe that continuum as well as the reduced
$Z$-pole annihilation are not capable of explaining the GC excess, but
the light Higgs pole scans through the required cross section.

In the right panel of Fig.~\ref{fig:bb} we show a zoomed-in version of
the Higgs peak. The interesting parameter regions for a combined fit
of the relic density with the GC excess will be given by
the solid relic density curves crossing the solid horizontal line and
the dashed GC lines crossing the dashed horizontal
line. As expected from the left panel, there are finely tuned regions
around the Higgs pole with today's velocity spectrum, which allow for
an explanation of the GC excess via a thermal relic through the process $\nne \nne \to b\bar{b}$.  
Decays of the light Higgs mediator to lighter fermions,
like tau leptons, are subleading because of the smaller Yukawa
coupling and the smaller color factor. Annihilation
through a $t$-channel stau generally results in an
annihilation rate which is too small.

\subsection*{$\chi \chi \to WW$}

\begin{figure}[t]
\includegraphics[width=0.495\textwidth]{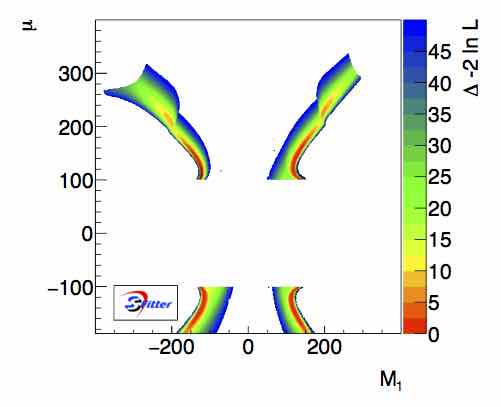}
\includegraphics[width=0.495\textwidth]{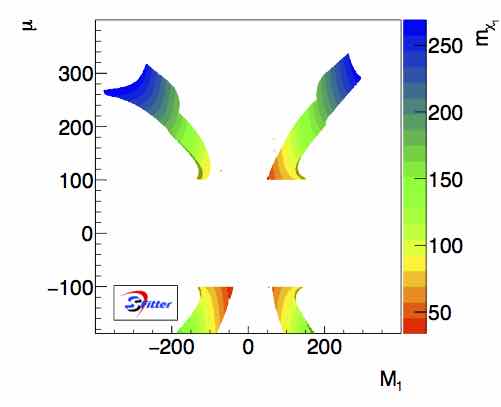}
\caption{Log-likelihood map (left) and corresponding LSP mass (right)
  based on the \fermilat photon spectrum for $M_2 = 700$~GeV and $\tan
  \beta = 45$, where $\nne \nne \to WW$ is a dominant
  annihilation channel. The heavy Higgses are decoupled to 1~TeV. The
  shaded dots are excluded by the \fermilat limits from dwarf spheroidal galaxies.}
\label{fig:mum1_ww}
\end{figure}

At slightly larger LSP masses, the dominant neutralino annihilation
channel is $\nne \nne \to WW$, mediated by a light chargino in the
$t$-channel (and chargino-neutralino co-annihilation for the relic
density).  Equation~\eqref{eq:w_coup} indicates that either wino or
higgsino LSP content enhances this annihilation rate. In
Fig.~\ref{fig:mum1_ww} we show the regions of the $M_1 - \mu$ plane
explaining the GC excess.  In this figure, we fix $M_2 = 700$~GeV,
implying that the LSP is a mixture of higgsino, coupling to
electroweak bosons, and bino. The preferred parameter range
compensates an increase in $|\mu|$ by an increase in $M_1$. This way
the sizeable higgsino content survives, while the neutralino mass
increases, as can be seen in the right panel of
Fig.~\ref{fig:mum1_ww}.  In the lower bands the allowed LSP masses
extend to $\mne{1} \approx 150$~GeV, without much decrease in the
log-likelihood.  The change in shape around $M_1 = |\mu| = 200$~GeV is
caused by the on-set of the annihilation to top pairs. The MSSM
parameter regions which allow for efficient annihilation in gauge
bosons are strongly correlated in $M_1$ and $\mu$, but not as tuned as
the light Higgs funnel region with its underlying pole
condition. Technically, this means that they are easy to identify in
a global fit.  In Fig.~\ref{fig:mum1_ww} we also indicate the
\fermilat limits from dwarf spheroidal galaxies~\cite{fermi_dwarfs} as black
dots. While these constraints are visible in the $M_1$ vs $\mu$ plane,
they do not significantly interfere with the best-fit regions from the
GC excess.

\subsection*{$\chi \chi \to t \bar{t}$}

\begin{figure}[t]
\includegraphics[width=0.495\textwidth]{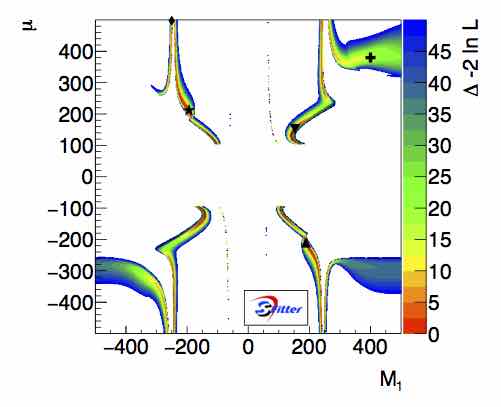}
\includegraphics[width=0.495\textwidth]{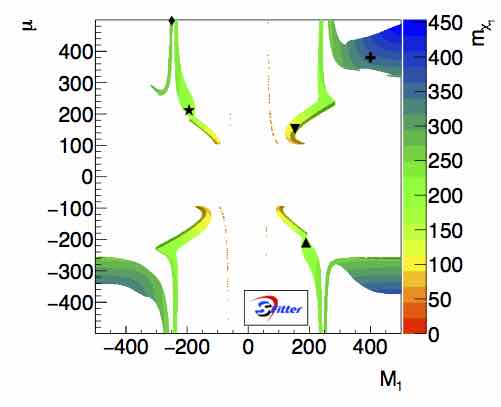}
\caption{Log-likelihood map (left) and corresponding LSP mass (right)
  based on the \fermilat photon spectrum for $M_2 = 700$~GeV, $\tan
  \beta = 3$, and $m_A = 500$~GeV, where we also observe the
  annihilation $\nne \nne \to t \bar{t}$.  The shaded dots are
  excluded by the \fermilat  limits from dwarf spheroidal galaxies.  The five symbols indicate
  local best-fitting parameter points.}
\label{fig:mum1_tt}
\end{figure}

Large annihilation cross sections for $\nne \nne \to t
\bar{t}$ can be accomplished by decreasing the heavy pseudoscalar mass
to $m_A = 500$~GeV and increasing the effective top Yukawa coupling by
choosing $\tan \beta=3$.  We show the allowed parameter range for
heavy winos, $M_2 = 700$~GeV, in Fig.~\ref{fig:mum1_tt}.  From
Fig.~\ref{fig:mum1_ww} we observe that for $\mne{1} > 175$~GeV the
annihilation into top pairs follows the $WW$ annihilation region in
the $M_1-\mu$ plane. We note that the $WW$ now behaves exactly the
same way, in spite of the lower choice of $\tan \beta$.

The primary difference is smaller $M_1$ values around $|\mu| =
200$~GeV. This increased bino fraction compensates the fact that the
underlying top Yukawa coupling is larger than the weak gauge coupling.
According to Fig.~\ref{fig:mum1_tt} the allowed mass range now extends
to $\mne{1} \gtrsim 200$~GeV. The main new feature for the reduced
value of $m_A = 500$~GeV is the peak towards large $\mu$ values for
$M_1 \approx 300$~GeV. The corresponding LSP mass is around 250~GeV,
close to the $A$-pole.  On the pole, annihilation is too efficient and
the preferred coupling is reduced by a smaller higgsino fraction in
the LSP. Beyond the pole, the allowed region extends to LSP masses
above 250~GeV, but with a reduced log-likelihood. If we choose larger
values of $\tan \beta$ the same structure remains, but the narrow pole
gets washed out into a wider band of dark matter masses.  The fact
that this large-$|M_1|$ regime does not appear in the upper left
corner of Fig.~\ref{fig:mum1_tt} is explained by the feature that at
tree-level this region of parameter space features $\mch{1} <
\mne{1}$, though this ordering will most likely be reversed by loop
corrections~\cite{mass_loops}.

\begin{figure}[t]
\centering
\includegraphics[width=0.6\textwidth]{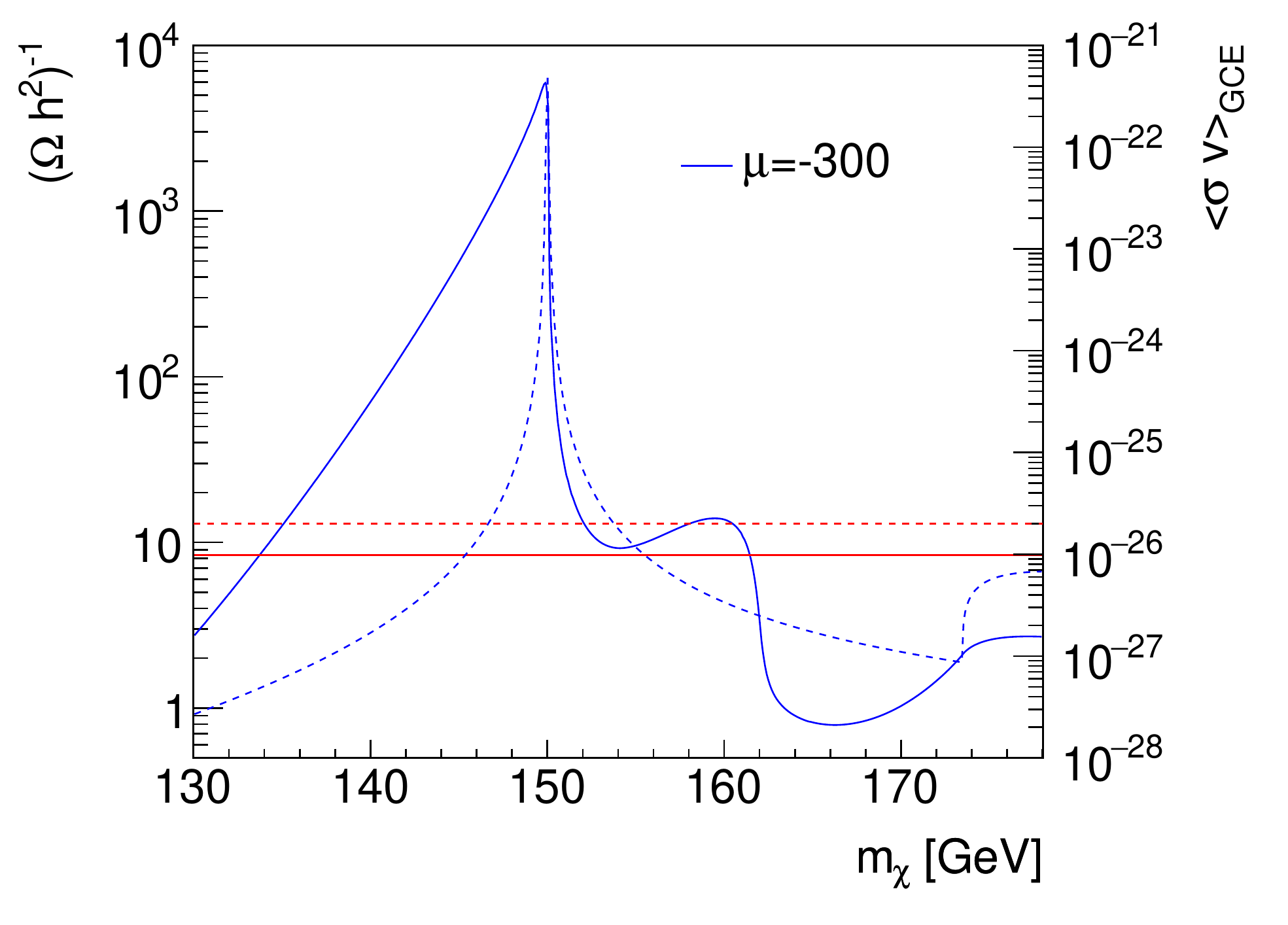}
\caption{Inverse relic density (dashed, left axis) and annihilation
  rate in the GC (solid, right axis) for an MSSM
  parameter point where the annihilation receives a contribution from
  $\nne \nne \to hh$.}
\label{fig:hh}
\end{figure}

\subsection*{$\chi \chi \to hh$}

In principle, for $\mne{1} > m_h$ the LSP can also annihilate to a
pair of SM-like Higgs bosons, $\nne \nne \to hh$.  While the
$t$-channel neutralino diagram will typically be overwhelmed by the
annihilation to weak bosons with the same $t$-channel mediator, an
$s$-channel mediator with $m_h \approx 2 m_h$ can dominate for small
$\tan \beta$. In Fig.~\ref{fig:hh} we show how the corresponding
effect for dark matter annihilation in the early universe (left axis)
and in the GC (right axis), similar to the $b\bar{b}$ case in
Fig.~\ref{fig:bb}. The LSP mass is varied through $M_1$, while $\mu =
-300$~GeV and $M_2 = 700$~GeV. The heavy Higgses are light, namely
$m_A = 300$~GeV and $m_H \approx 320$~GeV. The heavy Higgs' branching
ratio to a pair of light Higgses is $\br(H \to hh) =
30\%$~\cite{higgs_to_higgs}.  For comparably large velocities we see
how both $s$-channel mediators, $H$ and $A$, contribute through their
respective on-shell configuration.  In contrast, for the smaller
velocities associated with the \fermilat GC excess the CP-odd mediator
$A$ completely dominates, while the CP-even $H$ does not contribute
visibly. Because only the latter couples to two light Higgs bosons,
the annihilation to Higgs pairs leading to the GC excess is difficult
to realize in the MSSM. This outcome is different from the case of a
single-scalar Higgs portal model~\cite{portal}.  The increase we
observe in Fig.~\ref{fig:hh} for $\mne{1} > 170$ again shows the onset
of the annihilation into two tops.\bigskip

\begin{figure}[t]
\includegraphics[width=0.495\textwidth]{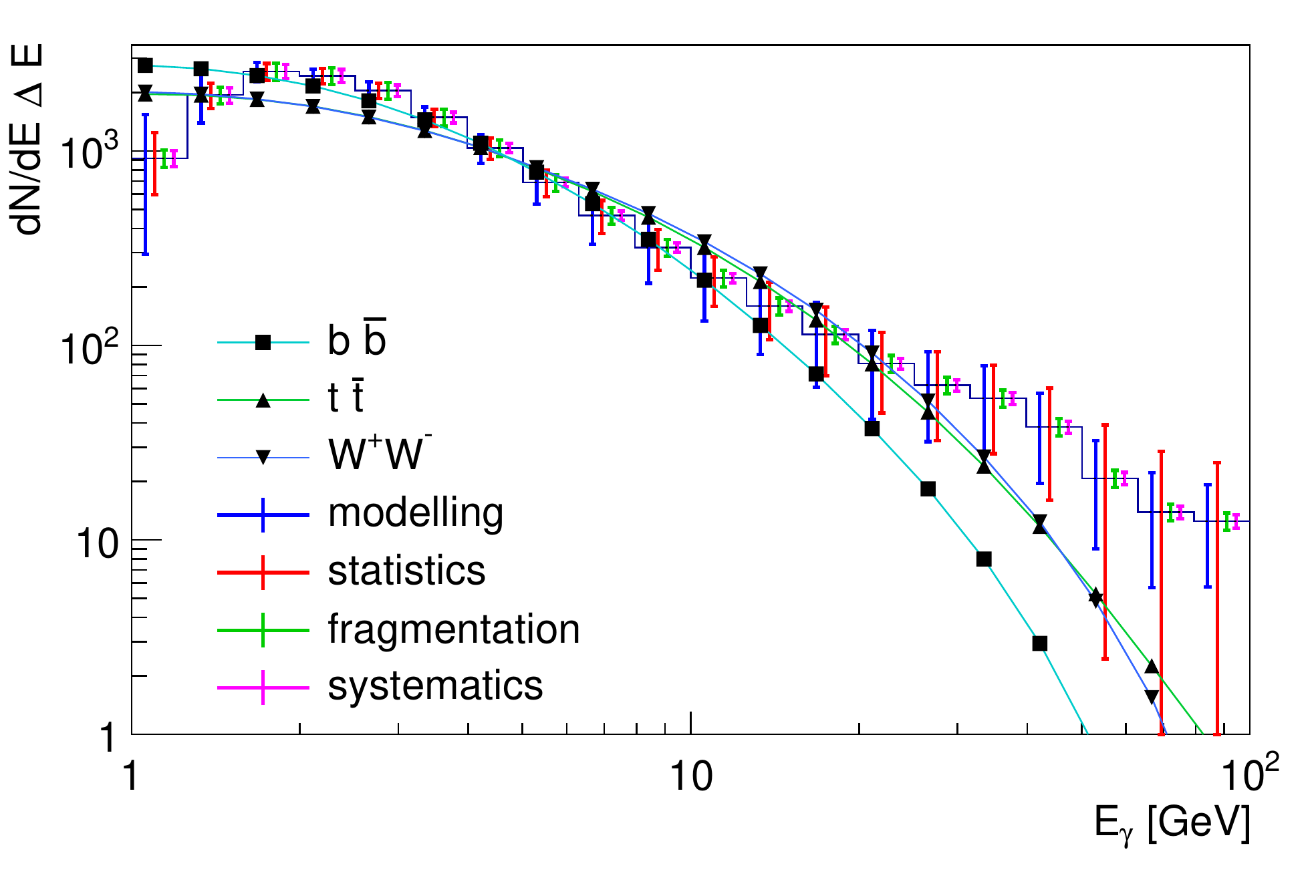}
\includegraphics[width=0.495\textwidth]{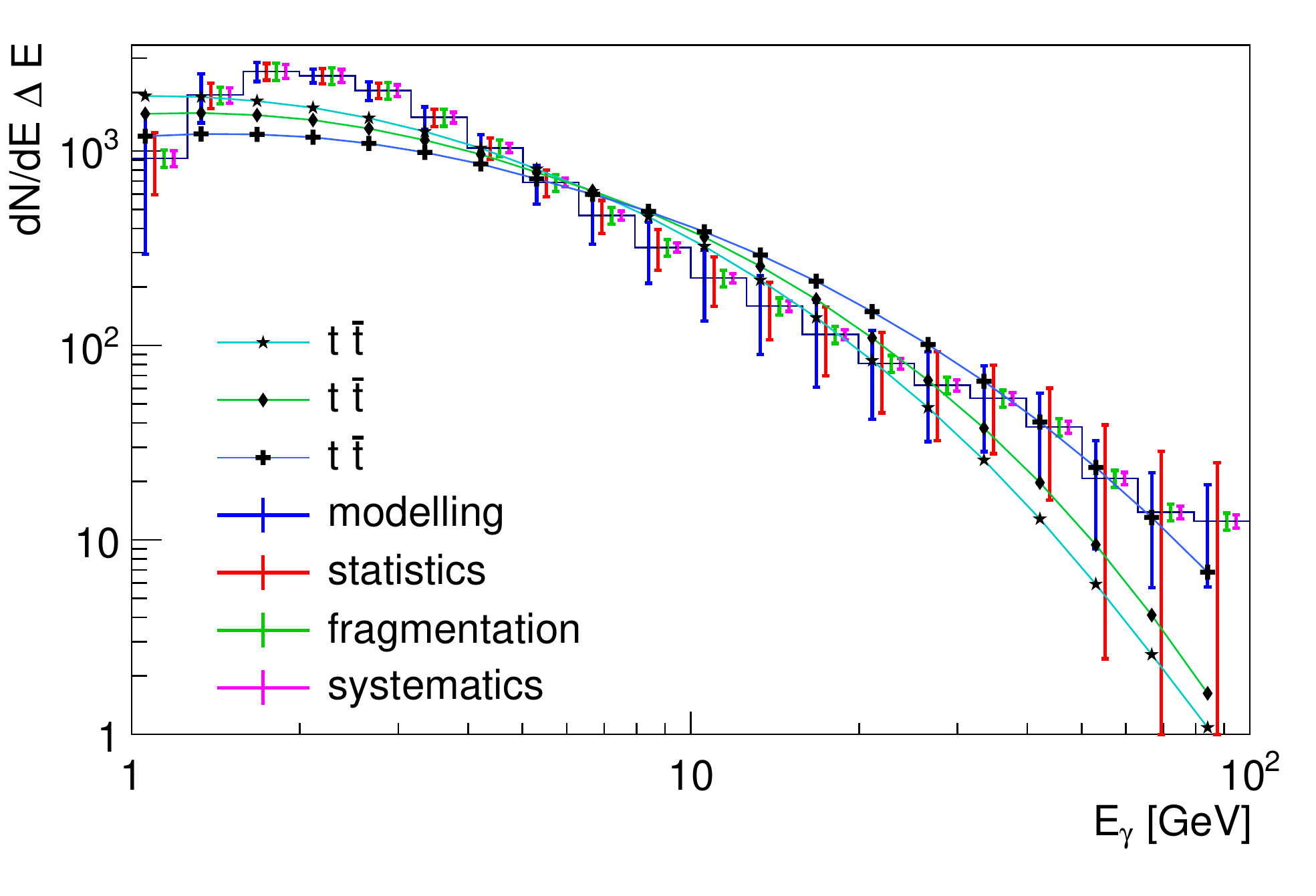}
\caption{Spectra for local best fitting MSSM parameter points,
  assuming dark matter annihilation dominantly to $b\bar{b}, WW,
  t\bar{t}$ (left) and for three different $t\bar{t}$ annihilation
  channels (right). The markers correspond to Fig.~\ref{fig:mum1_tt}, as indicated.}
\label{fig:spectra}
\end{figure}

Based on these example scenarios it is now clear that the GC excess can
be realized by the dominant annihilation channels
\begin{align}
\nne \nne \to b\bar{b}, WW, t\bar{t}
\end{align}
in more or less finely tuned parameter ranges of the MSSM. At this
level, the assumed value of $\tan \beta$ plays a role in how hard it
is to arrive at the correct light Higgs mass and how often the heavy
Higgses decay to up-type and down-type fermions. Annihilation to light
fermions like $b\bar{b}$ is realized through a finely tuned, resonant
$s$-channel mediator. In addition, the LSP can be a neutralino with
$\mne{1} = 100~...~350$~GeV with dominant annihilation to $WW$ and/or
$t\bar{t}$ pairs.  In Fig.~\ref{fig:spectra} we show a set of sample
energy spectra for different scenarios, defined as five local
best-fitting points in Fig.~\ref{fig:mum1_tt}. We overlay the
\fermilat spectrum shown in Fig.~\ref{fig:spectrum}. The three
scenarios with leading decays to $b\bar{b}$, $WW$, and $t\bar{t}$
shown in the left panel agree with the \fermilat results similarly
well. The lowest-energy and highest-energy bins cause the
largest problem in particular for a light LSP with Higgs funnel
annihilation into $b\bar{b}$ pairs. In the right panel of
Fig.~\ref{fig:spectra} we show three different parameter points, all
with a leading annihilation to $t\bar{t}$ pairs, and with LSP masses
$\mne{1} = 190, 255$, and 350~GeV. The over-all agreement with the
\fermilat spectrum gets slightly worse towards larger masses, leading
to a Gaussian-equivalent $\Delta \chi^2 = 18$ between the three
curves.

\section{MSSM analysis}
\label{sec:global}

\begin{table}[b!]
\vspace*{0.5cm}
  \begin{tabular}{l l l}
    \hline
    Measurement & Value \\
    \hline
    $m_h$ & $(125.09\pm 0.21_\text{stat} \pm 0.11_\text{syst}\pm3.0_\text{theo}  )~\gev$ & \cite{hmass,m_h} \\
    $\Omega_\chi h^2$ & $0.1188 \pm 0.0010_\text{stat} \pm 0.0120 _\text{theo}$ & \cite{hooperon_planck}\\
    $a_{\mu}$ & $(287 \pm 63_\text{exp}  \pm 49_\text{SM} \pm 20_\text{theo}   )\cdot 10 ^{-11}$ & \cite{Amu}\\
    $\br(B\rightarrow X_s \gamma)$ & $ (3.43 \pm 0.21 _\text{stat} \pm 0.07 _\text{syst}) \cdot 10^{-4}$ & \cite{XsGamma}\\
    $\br(B_s^0 \rightarrow \mu^+ \mu^-)~~~~~~~~~ $ & $ (3.2 \pm 1.4 _\text{stat} \pm 0.5 _\text{syst} \pm 0.2 _\text{theo}  )\cdot 10^{-9}$   & \cite{Bsmumu} \\
		$m_{\chi^+_1}$ & $>103~\gev$ & \cite{lep_constraints}\\
    \hline
  \end{tabular}
  \caption{Data used for the fit including their systematic,
    statistical, and theoretical uncertainties,  as appropriate.}
  \label{tab:data}
\end{table}

After understanding how different annihilation channels can be
realized in the MSSM we now perform a global analysis to determine the
range of MSSM parameter space which can best describe the GC
excess. This will be in the context of an LSP which makes up the
entirety of the dark matter and whose abundance is set by freeze-out
in a standard cosmology.  We impose the constraints shown in
Tab.~\ref{tab:data} by generating the MSSM spectrum and the
$B$-observables, and $(g -2)_\mu$ 
with \textsc{SuSpect3}~\cite{suspect}.  The relic density, indirect
detection rates, and direct detection rates are extracted from
\textsc{MicrOMEGAs}~\cite{micromegas}.  For $\mne{1} < 45$~GeV the
additional contribution to the invisible $Z$-width~\cite{z_inv} from
decays into pairs of LSPs plays a role~\cite{hooperon_sfitter}, but
in this analysis we do not have to take it into account.  The top mass
is fixed as an input, because the effect from the small range of
values consistent with collider measurements can be absorbed into
small shifts in the stop parameters. Limits from direct detection
experiments Xenon~\cite{xenon}, LUX~\cite{lux}, and
PandaX~\cite{pandax}, are only applied in the second part of this
section.\bigskip

\begin{figure}[t]
\includegraphics[width=0.495\textwidth]{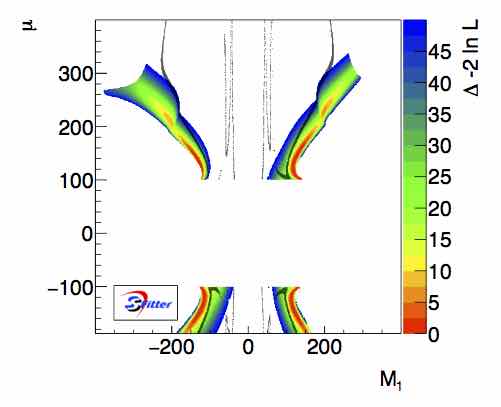}
\includegraphics[width=0.495\textwidth]{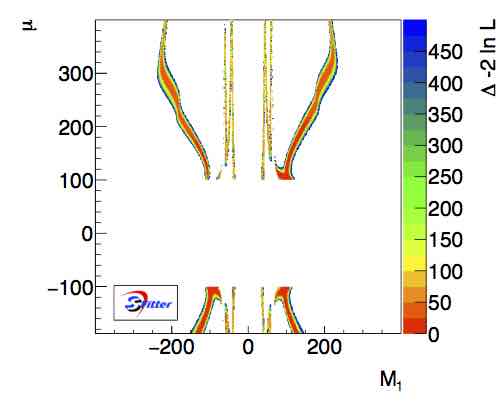} \\
\includegraphics[width=0.495\textwidth]{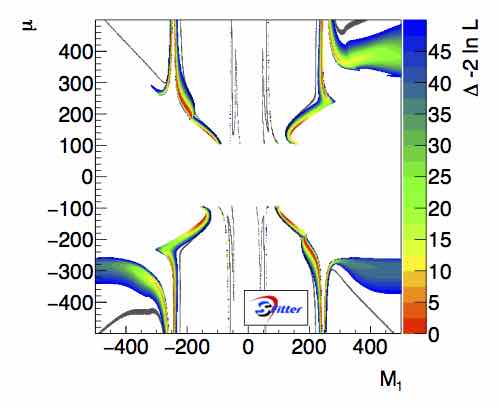}
\includegraphics[width=0.495\textwidth]{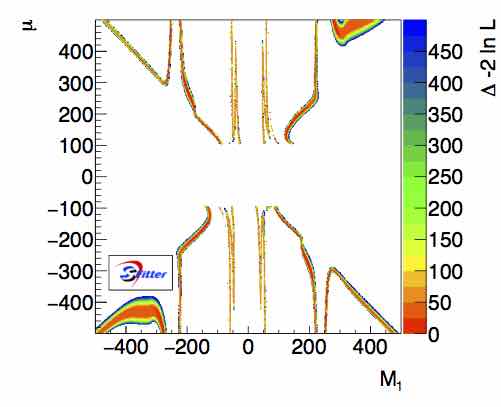}
\caption{Log-likelihood map including the \fermilat photon spectrum
  only (left) and in combination with the observed relic density, and
  other constraints (right) discussed in the text. We fix $M_2 =
  700$~GeV, $m_A = 1$~TeV, $\tan \beta=45$ (upper) or $m_A
  = 500$~GeV, $\tan \beta=3$ (lower), and vary $M_1$ and $\mu$. The
  black dots in the left panels are roughly compatible with the
  observed relic density.}
\label{fig:mum1_all}
\end{figure}

In the upper two panels of Fig.~\ref{fig:mum1_all} we show the allowed
parameter range in the bino and higgsino mass parameters, fixing the
wino mass to be essentially decoupled $M_2 = 700$~GeV and also decoupling
the heavy Higgses. The upper left panel mainly shows the $WW$ and
$b\bar{b}$ annihilation regions; in contrast to Fig.~\ref{fig:mum1_ww}
we also show the parameter points which give the correct relic
density $\Omega_\chi h^2$, quoted in Tab.~\ref{tab:data}. From
Fig.~\ref{fig:mum1_ww} we observe that the LSP masses in the $b\bar{b}$
scenario are very close to $\mne{1} = m_h/2$, while for the $WW$
scenario they extend to around $\mne{1} \approx 150$~GeV. As expected
from the similar underlying cross sections, the relic density and the
GC excess point to similar parameter regions, with
slightly larger $\mu$ for the relic density and hence smaller
annihilation cross sections $\langle \sigma v \rangle$.

In the upper right panel of Fig.~\ref{fig:mum1_all} we show the result
of a properly combined analysis of the GC excess and the measured
relic density.  Here, \textsc{SFitter} determines multidimensional
likelihood maps for the model parameter space. A set of Markov chains
selects points in the model space following a Breit--Wigner proposal
function. For each point we compute all considered observables and
determine a generalized $\chi^2$ value~\cite{sfitter,fittino}. For
this first analysis the likelihood map is 2-dimensional, covering
$M_1$ and $\mu$ over the range defined in the figures.

Because of the significantly smaller error bars, the
relic density measurement dominates the combined structures in the
$M_1$ vs $\mu$ parameter space. We observe three different
annihilation mechanisms: the vertical Higgs-pole $b\bar{b}$ peaks for
small $M_1$, the $WW$ region extending diagonally to $M_1 \approx
200$~GeV, and a continuum $t\bar{t}$ region for even larger values of
$M_1$.\bigskip

In the two lower panels of Fig.~\ref{fig:mum1_all} we show the same
parameters, but including a pseudo-scalar with
$m_A= 500$~GeV. The left panel illustrates the
$s$-channel annihilation regime and in particular above the $A$-pole
the relic density and the GC excess are difficult to reconcile. 
In the right panel we show how the combined fit follows the
relic density contours with its much smaller uncertainties. This
also implies that the asymmetry in the left panel with the missing
region at large negative $M_1$ and large positive $\mu$ re-appears in
the combined fit. Here the problem with $\mne{1} > \mch{1}$ does not
occur.  

\subsection*{Direct detection}
\label{sec:dd}

\begin{figure}[t]
\includegraphics[width=0.495\textwidth]{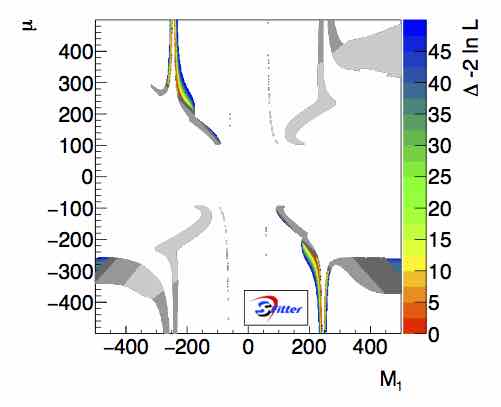}
\includegraphics[width=0.495\textwidth]{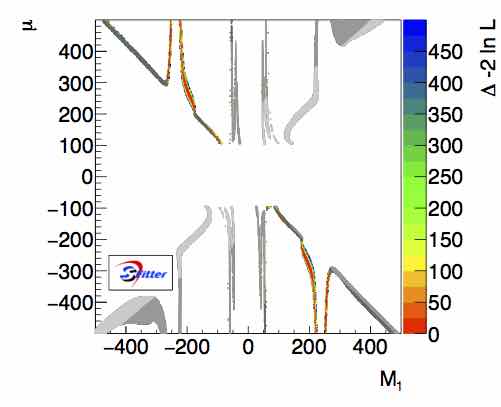}
\caption{Log-likelihood map including the GC excess,
  combined with direct detection constraints (left) and after adding
  the relic density, and other constraints discussed in the
  text (right). We fix $M_2 = 700$~GeV, $m_A = 500$~GeV,
  $\tan\beta = 3$, and vary $M_1$ and $\mu$. Different shades of gray
  indicate (from light to dark) the most recent exclusion limits from
  Xenon 100, PandaX and LUX.}
\label{fig:mum1_dd}
\end{figure}

An important, recently improved constraint comes from the direct
detection experiments probing coherent spin-independent scattering of
dark matter with a heavy nucleus.  In the left panel of
Fig.~\ref{fig:mum1_dd} we show the combination of the \fermilat GC
excess and different direct detection constraints, not including the
observed relic density and hence allowing for a non-standard
cosmology. Three shades indicate constraints from
Xenon100~\cite{xenon} (light), PandaX~\cite{pandax} (medium), and
LUX~\cite{lux} (dark). These constraints are included at face value
rather than in terms of a combined log-likelihood.  Instead of a
notoriously difficult error bar, we show three different rounds of
exclusion limits to illustrate the possible effect of weaker direct
detection constraints.  The remaining parameter points are colored
according to their combined \fermilat GC excess and indirect constraints
log-likelihood. All of the surviving parameter points rely on the
annihilation process $\nne \nne \to t\bar{t}$. The reason is that the
heavy (pseudo-)scalar mediator does not couple strongly to the
non-relativistic proton content, leaving the corresponding explanation
of the GC excess untouched.

For the right panel of Fig.~\ref{fig:mum1_dd} we combine the \fermilat
GC excess, direct detection constraints, the observed relic density,
and the other constraints shown in Tab.~\ref{tab:data}. As shown
before, the preferred regions in the $M_1 - \mu$ plane are now
slightly shifted and defined by the correct prediction of the relic
density.  With this modification, the $A$-funnel with an annihilation
to $t\bar{t}$ as well as a small range of points with the annihilation
signature $\nne \nne \to WW$ remain after direct detection
constraints. Throughout our analysis we only show log-likelihood
differences, the best-fit regions typically lead to a Gaussian
equivalent of $\chi^2/\text{d.o.f} \approx 1$.\bigskip

\begin{figure}[t]
\includegraphics[width=0.495\textwidth]{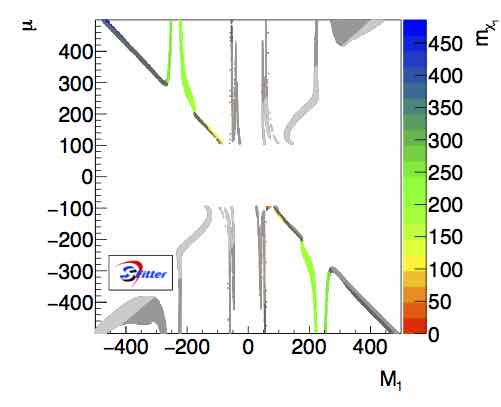}
\includegraphics[width=0.495\textwidth]{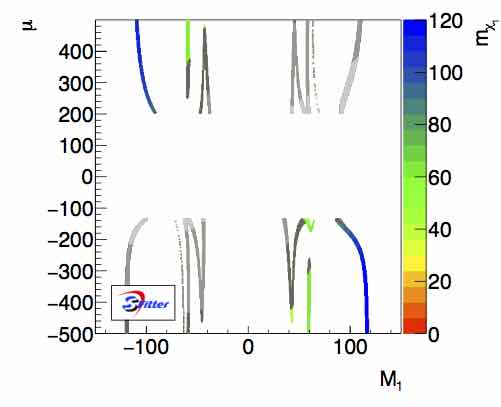}
\caption{Log-likelihood map including the \fermilat photon spectrum,
  direct detection constraints, the observed relic density, and other
  constraints discussed in the text for fixed $m_A = 500$~GeV and
  $M_2 = 700$~GeV, $\tan \beta=3$ (left) or $M_2 = 120$~GeV, $\tan
  \beta =7$ (right). Different shades of gray indicate (from light to
  dark) the most recent exclusion limits from Xenon100, PandaX and
  LUX.}
\label{fig:lowm2_dd}
\end{figure}

A key parameter is the mass of a dark
matter candidate which simultaneously explains the observed relic
density and the GC excess, and at the same time respects
all constraints in Tab.~\ref{tab:data} as well as those from direct detection experiments.
In Fig.~\ref{fig:lowm2_dd} we show all points with $\Delta (-2 \log L)
\lesssim 50$, colored according to the LSP mass $\mne{1}$.  In the
left panel we fix $M_2 = 700$ and $\tan \beta = 3$, as before. The low
value of $m_A = 500$~GeV opens a $t\bar{t}$ annihilation region
with $\mne{1} \approx 200$~GeV. In addition we see a few allowed
points with $\mne{1} \lesssim 100$~GeV in the $WW$ regime.

In the right panel of Fig.~\ref{fig:lowm2_dd} we fix $M_2=120$~GeV,
allowing for a significant wino fraction in the LSP. According to
Eq.\eqref{eq:w_coup} the wino content generally allows for a sizable
annihilation rate through a $t$-channel chargino, implying that the
LSP mass after requiring the annihilation rate matching the GC excess as well as the observed relic density will never exceed
120~GeV. On the other hand, a higgsino admixture can lead to lighter
valid dark matter candidates.  We again identify the very narrow
$h$-peak and the broader $Z$-peak. They define the allowed parameter
points with $\mne{1} \approx 45$~GeV and $\mne{1} \approx 63$~GeV.  In
addition, we see a non-resonant band of allowed points with $\mne{1} =100~...~120$~GeV, 
with an annihilation into $WW$ pairs. Annihilation
into a pair of top quarks is kinematically impossible.  Direct detection experiments
have a weaker impact because gaugino mixtures have smaller couplings to the light
Higgs.

In summary, we see that in particular for a mixed wino-higgsino LSP
all three annihilation channels $b\bar{b}, WW, t\bar{t}$ survive
current direct detection limits, but with a much reduced number of
allowed parameter points. With the next generation of direct detection
experiments it should be possible to probe these remaining MSSM
parameter points.

\subsection*{Global parameter study}
\label{sec:full}

Finally, we perform a global MSSM fit in
the neutralino/chargino parameter space. To assure the possibility of the heavy
Higgs funnel we fix $m_A = 500$~GeV and vary:
\begin{alignat}{7}
\abs{M_1} &< 500~\gev &\qqqquad
\abs{M_2} &< 700~\gev &\qqqquad 
\abs{\mu} &< 500~\gev \notag \\
\abs{A_t} &< 7~\tev & 
\tan \beta &= 2~...~45 \; .
\end{alignat}
The remaining parameters, including squark masses, slepton masses, and
trilinear couplings, are decoupled at 4~TeV. This choice allows for
points interpolating between the two scenarios shown in
Fig.~\ref{fig:lowm2_dd}: bino-higgsino dark matter and
wino-higgsino dark matter.  In addition, the simultaneous variation of
$\tan \beta$ and $A_t$ ensures that for any value of $\tan \beta$ we
can generate the correct light Higgs mass while at the same time
scanning the bottom Yukawa coupling or the width of the heavy
Higgses.\bigskip

\begin{figure}[t]
\includegraphics[width=0.325\textwidth]{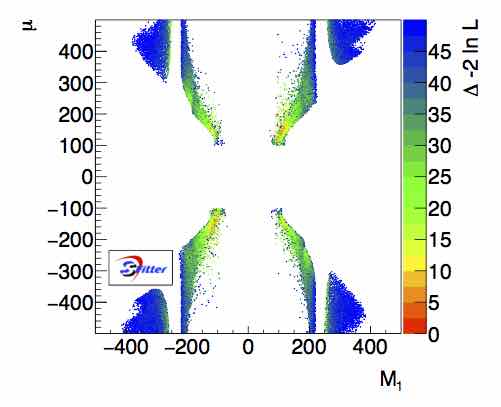}
\includegraphics[width=0.325\textwidth]{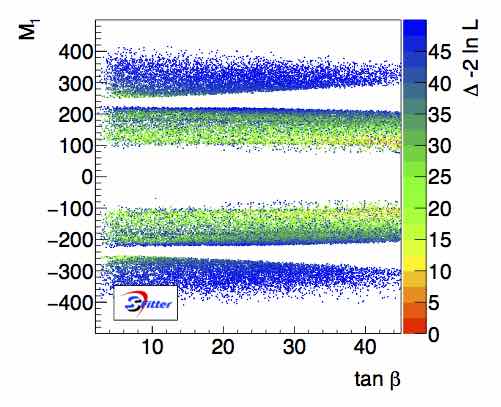} 
\includegraphics[width=0.325\textwidth]{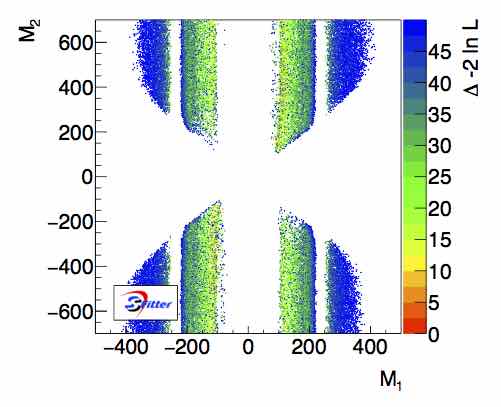} \\
\includegraphics[width=0.325\textwidth]{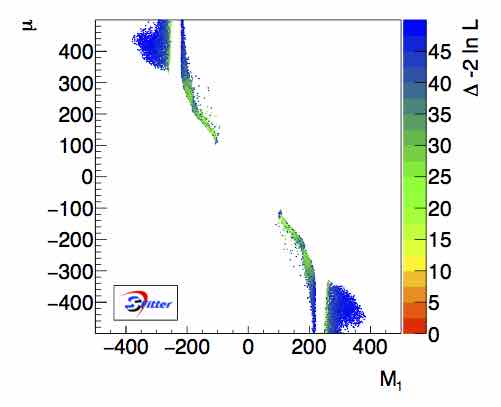}
\includegraphics[width=0.325\textwidth]{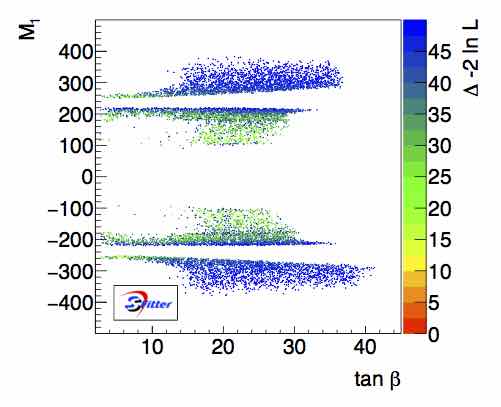}
\includegraphics[width=0.325\textwidth]{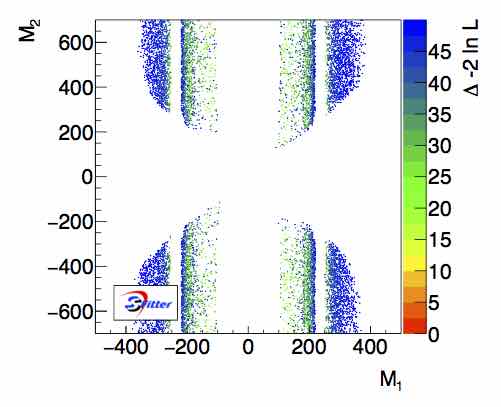} 
\caption{Log-likelihood map for the combined \textsc{Sfitter} analysis
  of the \fermilat photon spectrum, the observed relic density, and
  other constraints with (upper) and without (lower) including the LUX
  direct detection bounds.}
\label{fig:multi}
\end{figure}

In the upper panels of Fig.~\ref{fig:multi} we show the result of a
global analysis taking into account all constraints defined in
Tab.~\ref{tab:data}, but not including direct detection bounds. For
example the $\mu - M_1$ plane is now shown as a profile likelihood
after projecting our the remaining model parameters. In general, this
leads to a broadening of all features discussed before.  We still see
the usual narrow regions corresponding to the annihilation channels
$\nne \nne \to WW$ and $t\bar{t}$. In addition, broader structures for
large $|\mu| \sim |M_1|$ are generated by an the $\tan \beta$-enhanced
annihilation $\nne \nne \to A \to b\bar{b}$. They are much wider than
all other structures because the heavy Higgs width scales with $\tan^2
\beta$. In the second upper panel we observe that $\tan \beta$ has
hardly any global effect on the annihilation rate, both for the GC
excess and for the observed relic density. Towards large $\tan \beta$
we see how the low-$M_1$ scenarios reach a better agreement with data,
and how the width of the pseudoscalar Higgs with $m_A = 500$~GeV
increases.  Finally, in the right panel we observe a strong
correlation between $M_2$ and $M_1$, similar to the first panel, but
with more washed-out structures in the profile likelihood.  The
$Z$-funnel and $h$-funnel are not resolved by the usual global
analysis, and do not appear.  From the previous discussion, it is
clear that they are viable in the absence of direct detection
constraints.

In the lower panels of Fig.~\ref{fig:multi} we add the LUX direct
detection constraints.  All general structures in the $\mu - M_1$
plane, corresponding to the different decay channels, survive. An
independent sign change in $\mu$ and $M_1$ is no longer possible
because of the large degree of fine-tuning. The main difference
between this global result and the previous, two-dimensional analysis
is that for large $\mu \sim -M_1$ the pseudoscalar Higgs funnel
mediates an annihilation to $b\bar{b}$ pairs at large $\tan \beta$.

Another new feature in the global fit is an allowed higgsino LSP
region for $M_1 = 100~...~150$~GeV and $\tan \beta = 15~...~25$. It
corresponds to a combined annihilation to $WW$ and $ZZ$
pairs. Following Eq.\eqref{eq:w_coup} and Eq.\eqref{eq:z_coup} both,
the $\nne-\cpme-W$ and $\nne-\nne-Z$ couplings increase for large
$\tan \beta$. This way they lead to an efficient annihilation, but are
also ruled out by direct detection constraints. When we reduce $\tan
\beta \to 1$, the $\nne-\cpme-W$ coupling approaches a finite value,
while the $\nne-\nne-Z$ vanishes. 

\section{Outlook}

Based on a realistic estimate of the different sources of uncertainty
we have shown that the lightest neutralino in the MSSM can explain the
\fermilat GC excess. The different annihilation channels
$\nne \nne \to b\bar{b}$, $WW$, and $t\bar{t}$ define the
corresponding LSPs with increasing masses. The annihilation channel
$\nne \nne \to hh$ does not work in the MSSM, because of the velocity
suppression of the CP-even heavy Higgs funnel. Nevertheless, viable explanations of the GC excess in the MSSM
can annihilate to a wide range of Standard Model states and
cover a mass range from 45~GeV to well above 250~GeV.

If one demands that the LSP is a standard thermal relic, the preferred
regions of parameter space slightly shift. The typical width of the
structures in parameter space decreases significantly, corresponding
to the small uncertainties from the {\it Planck} fits. Consequently,
the the allowed region of a combined \textsc{SFitter} analysis follows
the patterns of the correct relic density. The best-fit region is
again defined by the $b\bar{b}$, $WW$, and $t\bar{t}$ annihilation
channels; it extends to LSP masses up to 300~GeV, in particular in
combination with a pseudoscalar heavy Higgs mass around 500~GeV.  In
addition, we confirm two more features in the MSSM parameter
space. First, a $\tan \beta$-enhanced annihilation of heavy
neutralinos to $b\bar{b}$ pairs can be mediated by the pseudoscalar
Higgs in complete analogy to the top quark case.  Second, the
different scaling of the neutral current and charged current couplings
of the neutralino/chargino sector opens an allowed wino region for
intermediate $\tan \beta$.

Finally, when we apply the full set of limits, the direct detection
constraint cuts deeply into the allowed parameter space. Nevertheless,
for a mixed wino-higgsino LSP all three annihilation channels with
their corresponding regions of parameter space survive. Most notably,
a heavy neutralino annihilating to top or bottom pairs remains largely
intact. Ignoring the relic density constraint and only considering the
GC excess combined with direct detection constraints does not improve
the situation qualitatively.  All of our preferred regions of
parameter space should be covered by the next generation of direct
detection experiments.

\begin{center}
\textit{Acknowledgments}
\end{center}

AB would like to thank the Heidelberg Graduate School for Fundamental
Physics for her PhD funding and the DFG Graduiertenkolleg
\textsl{Particle Physics beyond the Standard Model} (GK1940).  TP and
AB acknowledge the support by the DFG Forschergruppe \textsl{New
  Physics at the LHC} (FOR2238).  The work of SM is supported in part
by US department of energy grant DE-SC0014431.  The work of TMPT is
supported in part by US National Science Foundation grant PHY-1620638.


\end{fmffile}
\end{document}